\documentclass[twocolumn,superscriptaddress,secnumarabic,aps,pra,nobibnotes,groupedaddress,numerical]{revtex4-1}

\usepackage{graphicx}
\usepackage[squaren]{SIunits}
\usepackage{verbatim}
\usepackage{mathrsfs}
\usepackage{comment}
\usepackage{braket}
\usepackage{bbold}
\usepackage{amsmath}
\usepackage{hyperref}
\usepackage[normalem]{ulem}
\usepackage{xcolor}
\usepackage{times}
\usepackage{amsmath}

\begin{document}

\title{Emergent light crystal from frustration and pump engineering}

\author{Matteo Biondi}
\affiliation{Institute for Theoretical Physics, ETH Zurich, 8093 Z\"urich, Switzerland}
\author{Gianni Blatter}
\affiliation{Institute for Theoretical Physics, ETH Zurich, 8093 Z\"urich, Switzerland}
\author{Sebastian Schmidt}
\affiliation{Institute for Theoretical Physics, ETH Zurich, 8093 Z\"urich, Switzerland}
\pacs{XXX}

\begin{abstract}
We demonstrate how pump engineering drives the emergence of frustration-induced
quasi-long-range order in a low-dimensional photonic cavity array. We consider 
a Lieb chain of nonlinear cavities as described by the Bose-Hubbard
model and featuring a photonic flat band in the single-particle spectrum.
Incoherent pumping of the Lieb lattice leads to a photonic density-wave which
manifests an algebraic decay of correlations with twice the period of the
lattice unit cell. This work opens up new directions for the emergence of
strongly-correlated phases in quantum optical frustrated systems through pump
design.
\end{abstract}

\maketitle

Many-body physics with light is inspired by the combined opportunities offered
by quantum optics and condensed matter physics. Superconducting circuits
\cite{schmidt2013*2} and exciton-polaritons in semiconducting micro-cavities
\cite{carusotto2013} provide platforms for exploring strongly correlated
photons with light-matter induced interactions \cite{hartmann2016,noh2017}. The inherently
nonequilibrium nature of photonic systems adds further opportunities, since
the dynamics is not generated by the Hamiltonian alone but rather involves a
Liouvillian superoperator acting on the system's density-matrix that
incorporates unitary as well as dissipative dynamics. Drive and dissipation
then can be tailored to target a desired subspace governing the long-time
system dynamics. For instance, reservoir engineering \cite{cirac1993} has been
demonstrated within various quantum optical settings, e.g., trapped ions 
\cite{kienzler2015} and can be utilized in quantum
computation \cite{verstraete2009} or for generating topological phases
\cite{diehl2011}. Alternatively, exotic many-body states of nonequilibrium
systems can be engineered under continuous driving, as recently 
implemented for entangled steady-states \cite{lin2013,shankar2013} 
or proposed for fractional quantum Hall states of light \cite{kapit2014,umucalilar2017}. 
Here, we demonstrate that pump engineering leads to the emergence of
quasi-long-range order in a frustrated photonic lattice. 

While frustration in equilibrium systems, e.g., in spin lattices
\cite{castelnovo2008,villain1980,bergman2007}, Josephson junction arrays 
\cite{pannetier1984,vidal1998} and ultracold atoms \cite{huber2010,apaya2010,tovmasyan2013} has been well explored, 
only few pioneering efforts have been undertaken in quantum optics \cite{schmidt2016}. E.g.,
experiments on frustrated photonic lattices have demonstrated single-particle
interference \cite{vicencio2015,mukherjee2015}, all-optical logical operation
using flat band eigenstates \cite{real2017}, and exciton-polariton
condensation on a flat band with disorder-induced decoherence
\cite{baboux2016}. Recently, it has been proposed that a coherently driven
frustrated lattice develops an incompressible steady-state \cite{biondi2015}
characterized by exponentially decaying correlations. Here, we demonstrate
that pump engineering allows for the emergence of truly crystalline
quasi-long-range order.

\begin{figure*}[t!]
\includegraphics[width=0.785\textwidth]{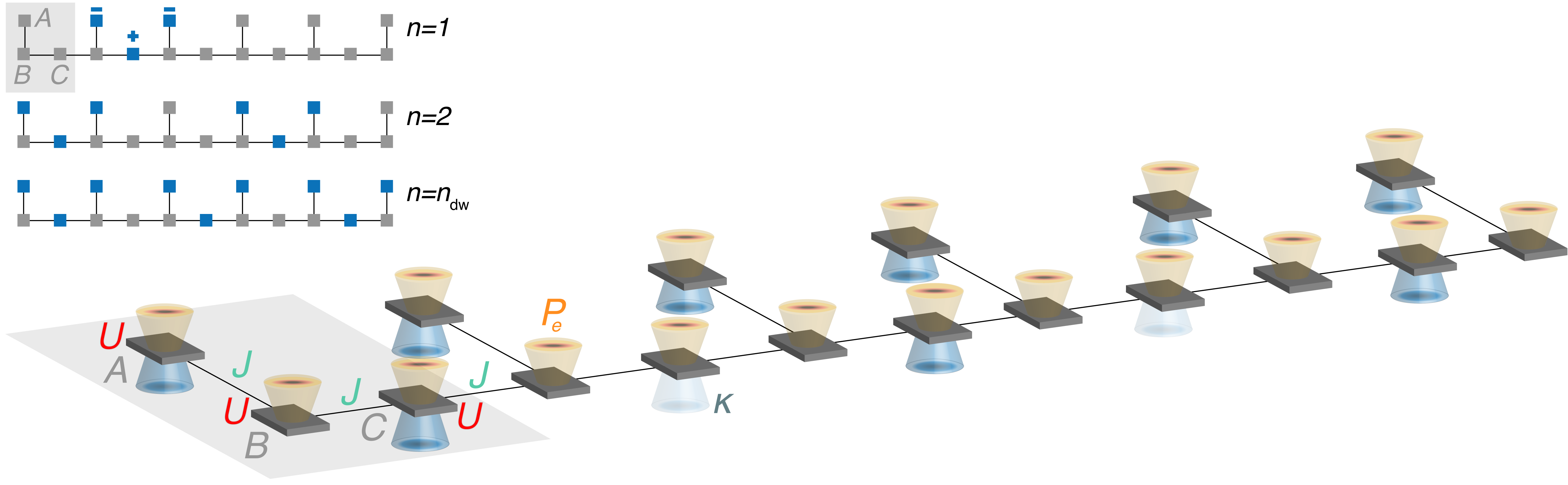}
\caption{Schematic view of the pumped and dissipative
quasi-one-dimensional Lieb lattice.  The lattice has a three-site unit cell
with basis sites $A, B, C$ (grey).  Here, the lattice displays $N=5$ cells and
terminates with $A$ and $C$ sites for symmetry. Photons can hop between
neighboring sites with amplitude $J$ (green) and experience an on-site
repulsion $U$ (red), as described by the Bose-Hubbard Hamiltonian, see
equation \eqref{hammodel}.  The Lieb lattice hosts a set of single-particle
(photon number $n=1$) plaquette states localized on one $C$ site and two
neighboring $A$ sites (top-left inset); these originate from destructive
quantum interference of amplitudes at site $A$, see equation
\eqref{lambda_state}. Product states of non-overlapping plaquettes are
many-body eigenstates of the Hamiltonian with zero interaction energy.  The
lattice is pumped homogeneously with strength $P_e$ (orange light cones) and loses
photons to the environment with a rate $\kappa$.  The nonequilibrium dynamics
is described by the master equation \eqref{me}. Due to the energy-dependent
spectral density of the pump, the steady-state is a density-wave where only
every second $C$ site is occupied (blue light cones) giving rise to
frustration-induced quasi-long-range order, see equation
\eqref{eq:correlator}.
\hfill \label{fig1}}
\end{figure*}

\begin{figure*}[t!]
\includegraphics[width=1\textwidth]{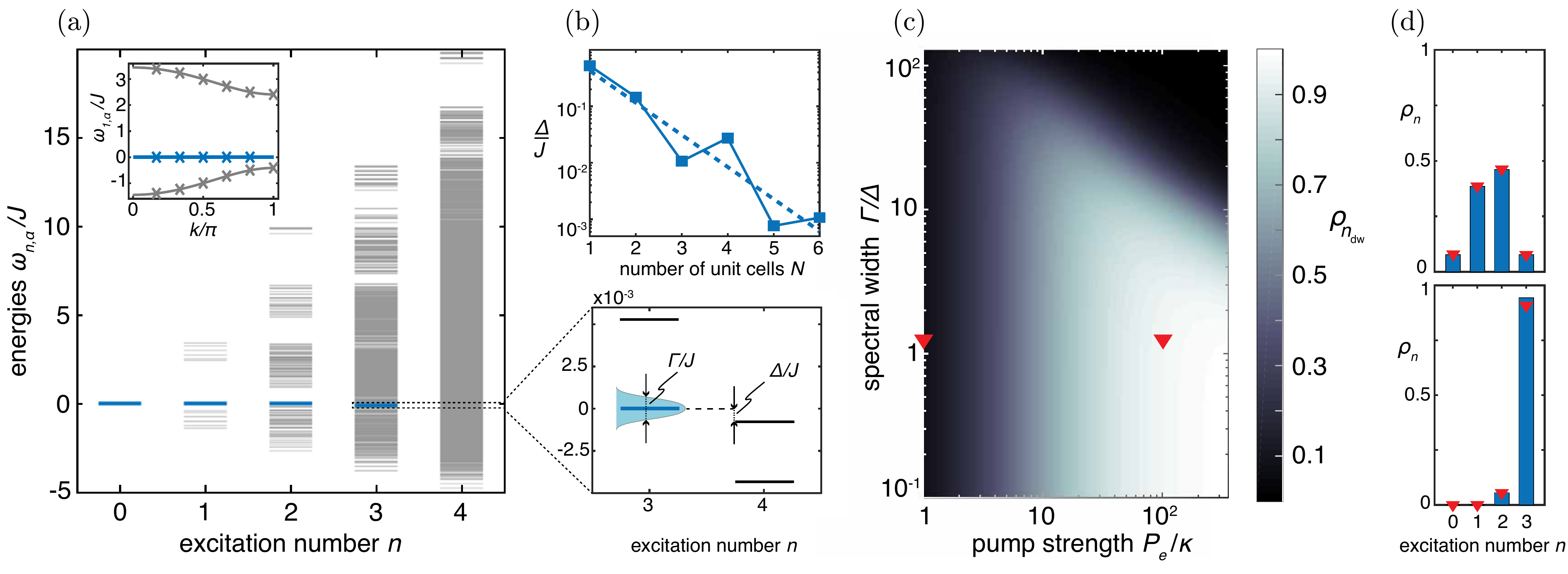}
\caption{(a)
Energy levels $\omega_{n,\alpha}/J$ of the quasi-1D Lieb lattice
\eqref{hammodel} with $N\!=\!5$ unit cells, i.e., $3N\!+\!2$ sites, as a
function of excitation (photon) number $n$, while the index $\alpha$ labels energies at 
fixed $n$. The levels $\omega_{n,\alpha}$ are scaled to the hopping amplitude $J$
and have been obtained via exact diagonalization (see SM).  Frustration
in the Lieb chain leads to a set of single-particle ($n\!=\!1$) plaquette
states at zero-energy, see equation \eqref{lambda_state}; product states of
non-overlapping plaquettes are also zero-energy eigenstates with $n>1$ up to
the density-wave state, see equation \eqref{dw_state}, with $n_{\rm
dw}\!=\!(N\!+\!1)/2$, i.e., $n_{\rm dw}\!=\!3$ in (a).  This manifold is
marked by the blue lines (including the vacuum with $n\!=\!0$). The top-left
inset shows the single-particle levels $\omega_{n=1,\alpha}$ in momentum space
(crosses), including lines for large $N$ (note the zero-energy flat band). The
band structure asymmetry is due to a finite offset $\omega_{\scriptscriptstyle
A}\!\neq\!\omega_{\scriptscriptstyle B}$. (b) The zero-energy manifold
is pumped (see text) using a Lorentzian spectral density.  Occupation of other
levels is suppressed if the width $\Gamma$ is smaller than the minimal
excitation gap $\Delta$ to higher bands with $n\!=\!n_{\rm dw}\!+\!1$ (bottom
panel). The gap decreases with system size, see top panel in (b).  The
dashed line is an exponential fit, $\Delta/J\propto\exp{(-\gamma N)}$,
$\gamma=1.31\pm0.67$. (c) Occupation of the density-wave $\rho_{n_{\rm
dw}}$ from the numerical solution of the master equation \eqref{me} as a
function of the Lorentzian width $\Gamma/\Delta$ and pump strength
$P_e/\kappa$ for the lattice with $N\!=\!5$ unit cells as in (a). For
small $\Gamma$ and large $P_e$ the steady-state of the Lieb chain is a
quasi-pure density-wave. (d) Populations within the flat band manifold,
as obtained from the analytical solution of the master equation \eqref{me} in
the projected Hilbert space \eqref{me_analytic} (bars), versus numerical
results (symbols), for $\Gamma/\Delta\!=\!1.3$ and $P_e/\kappa\!=\!1$ (top)
and $P_e/\kappa\!=\!10^2$ (bottom, cf.\ points marked by triangles in (c))
showing excellent agreement supporting the validity of the analytic study.
The other parameters are chosen as, $U/J\!=\!2.5$,
$\kappa/\Gamma\!=\!10^{-3}$, $\omega_{\scriptscriptstyle B}\!=\!\omega_{\scriptscriptstyle
A}\!+\!2J$.  \hfill \label{fig2}}
\end{figure*}

We consider a quasi-one-dimensional (1D) Lieb lattice \cite{lieb1989} of coupled nonlinear
cavities (see Fig.~\ref{fig1}) described by the Bose-Hubbard model with
Hamiltonian ($\hbar = 1$)
\begin{eqnarray}
\begin{split}
\label{hammodel}
   H & = \sum_{j}\sum_{\scriptscriptstyle I=A,B,C}
   \left[\,\omega_{\scriptscriptstyle I} p^\dagger_{j, \scriptscriptstyle I}
   p_{j,\scriptscriptstyle I} + U p^\dagger_{j, \scriptscriptstyle I}
   p^\dagger_{j, \scriptscriptstyle I}p_{j,\scriptscriptstyle I}
   p_{j,\scriptscriptstyle I}\,\right] \\ 
   & + J \sum_{j} \left[ p^\dagger_{j,\scriptscriptstyle B}\!
   \left(p_{j,\scriptscriptstyle A} + p_{j,\scriptscriptstyle C}\right) 
   + p^\dagger_{j+1,\scriptscriptstyle B}p_{j,\scriptscriptstyle C}  
   + \text{H.c.}\right]
\end{split}
\end{eqnarray}
Here, the bosonic operators $p^\dagger_{j,\scriptscriptstyle I}$ create a
photon at site $I = A, B, C$ in unit cell $j$ of the Lieb lattice.  The first
line in the Hamiltonian \eqref{hammodel} contains the on-site energies
$\omega_{\scriptscriptstyle I}$ and Kerr-type interaction $U$, while the
second line describes photon hopping between nearest-neighboring sites with a
rate $J$. In the following, we assume $\omega_{\scriptscriptstyle A} =
\omega_{\scriptscriptstyle C}$ and $\omega_{\scriptscriptstyle A} \neq
\omega_{\scriptscriptstyle B}$.  We also set $\omega_{\scriptscriptstyle A} =
0$ for convenience, while the precise value chosen for
$\omega_{\scriptscriptstyle B}$ does not affect the main results of this
paper.  The Lieb lattice is originally two-dimensional and is formally
obtained from a square lattice by adding a new site at the mid-points of all
bonds; when $\omega_{\scriptscriptstyle A} = \omega_{\scriptscriptstyle C}$
this \emph{decorating procedure} ensures that the band
structure of the Lieb lattice exhibits a flat band with energy
$\omega_{\scriptscriptstyle B} = 0$, i.e., a dispersionless band in the entire
Brillouin zone which arises from quantum interference \cite{tasaki1992}.  The
lattice considered in our manuscript is a quasi-1D cut of the 2D Lieb lattice
that still exhibits a flat band. The flat band is associated
with a set of degenerate, single-particle \emph{plaquette} states
$\ket{V_j}$ localized on a compact portion of the lattice,
\begin{eqnarray}
\label{lambda_state}
   \ket{V_j} = \frac{1}{\sqrt{3}}\big(p^\dagger_{j,\scriptscriptstyle C} 
   - p^\dagger_{j,\scriptscriptstyle A} 
   - p^\dagger_{j+1,\scriptscriptstyle A}\big)\ket{\text{vac}}
\end{eqnarray}
With three sites per unit cell of the quasi-1D Lieb lattice, the
single-particle band structure exhibits two more bands with a finite
dispersion (see inset in Fig.~\ref{fig2}(a)).

The peculiarity of the flat band manifests itself in the many-body problem:
starting from the plaquette states \eqref{lambda_state}, exact many-body
eigenstates of the Hamiltonian \eqref{hammodel} with zero-energy can be
constructed from product states of non-overlapping plaquettes, e.g., the
two-photon states $\ket{V_1}\ket{V_3}, \ket{V_1}
\ket{V_4}$, the three-photon states $\ket{V_1} \ket{V_3}
\ket{V_5}$ etc., up to the maximally filled density-wave state
\begin{eqnarray}
\label{dw_state}
   \ket{n_\text{dw}} = \prod_{j=1}^{n_{\rm dw}} \ket{V_{2j-1}}
\end{eqnarray}
whose photon number depends only on the geometry, i.e., $n_{\rm dw} = (N+1)/2$
for an odd number of unit cells $N$, see inset in Fig.~\ref{fig1}.  The filling is
given by $\nu_{\rm dw} = n_{\rm dw}/N_s = 1/6 + \mathcal{O}(1/N)$, where
$N_s=3N\!+\!2$ is the number of sites. Products of plaquettes with filling
higher than $\nu_{\rm dw}$ must double occupy at least one site
\cite{huber2010} and are therefore gapped from the zero-energy manifold by a
finite interaction energy $\propto \Delta$, see Fig.~\ref{fig2}(b).  The set
of flat band states forms a zero-energy manifold with a large degeneracy
$\sum_{n=0} ^{n_{\rm dw}} D_n =  F_{2n_{\rm dw} + 1}$, where $D_n = \binom{2
n_{\rm dw} - n}{n}$ is the degeneracy of the photon number sector $n$ and
$F_n$ is the Fibonacci number, $F_n \sim [(1 + \sqrt{5})/2]^n/\sqrt{5}$ for
large $n$.  Note, that the single-particle plaquette states
\eqref{lambda_state} are linearly independent but not orthogonal since
neighboring plaquettes share one site; we thus perform a Gram-Schmidt
orthogonalization of the flat-band manifold in each excitation sector
separately, since states with different excitation numbers are orthogonal, see
Supplemental Material (SM).

The density-wave \eqref{dw_state} exhibits strong density correlations with a
period twice that of the lattice (period doubling) since only every second $C$
site is occupied, thus representing a truly crystalline state. 
There have been much efforts lately towards 
the implementation of strongly-correlated states in frustrated system, 
originally in the context of ultracold atoms \cite{huber2010,taie2015} and
more recently in quantum optical systems
\cite{vicencio2015,mukherjee2015,real2017,biondi2015,baboux2016}. In the former
(equilibrium) case, the task is highly nontrivial since the flat band is not
the lowest energy band and a fully adiabatic transfer of the condensate would
be required \cite{taie2015}. Alternatively, in frustrated lattices of similar
kind, e.g., the sawtooth chain or the kagome lattice where the flat band is
the topmost energy band, equilibrium frustration can be investigated only
through band inversion, e.g, by implementing complex-valued tunnelling
constants \cite{aidelsburger2015}. In nonequilibrium photonic
systems, coherent driving allows to populate the flat band manifold
\cite{vicencio2015,mukherjee2015} but leads to a mixed steady-state with weak
occupation of the density-wave \cite{biondi2015}.

Here, we propose a radically different route which allows to engineer a
quasi-pure density-wave exhibiting an algebraic decay of photonic
density-density correlations.  To this end, the Lieb chain is incoherently
coupled to an external pump reservoir exhibiting a Lorentzian-shaped spectral
density centered at zero energy. This coloured environment can be engineered,
e.g.,  in circuit QED using microwave filters \cite{hoffman2011} or ancilla
quantum systems \cite{lebreuilly2016}. The dynamics of the system is then
described by a time-convolutionless master equation in the Born approximation
\cite{petruccione2002,hoffman2011,lebreuilly2016} for the lattice density
matrix
\begin{equation}
   \dot{\rho} = i[\rho,H] + (\kappa/2)\sum_{j,\scriptscriptstyle I}
   \mathcal{L}_{j,\scriptscriptstyle I}^\downarrow[\rho] 
   + (P_e/2)\sum_{j,\scriptscriptstyle I}
   \mathcal{L}_{j,\scriptscriptstyle I}^\uparrow[\rho].
\label{me}
\end{equation}
Here, we have introduced the Lindblad dissipators
$\mathcal{L}_{j,\scriptscriptstyle I}^\downarrow[\rho] =
2p_{j,\scriptscriptstyle I} \rho\, p_{j,\scriptscriptstyle I}^\dagger -
p_{j,\scriptscriptstyle I}^\dagger p_{j,\scriptscriptstyle I} \rho-\rho
p_{j,\scriptscriptstyle I}^\dagger p_{j,\scriptscriptstyle I}$ and the
generalized superoperators $\mathcal{L}_{j,\scriptscriptstyle
I}^\uparrow[\rho] = p_{j,\scriptscriptstyle I}^\dagger \rho\,
\tilde{p}_{j,\scriptscriptstyle I} - \rho\, \tilde{p}_{j,\scriptscriptstyle
I}p_{j,\scriptscriptstyle I}^\dagger + \text{H.c.}$ The first term in equation
\eqref{me} describes the coherent evolution under the lattice Hamiltonian $H$,
the second accounts for photon loss from each site with a rate $\kappa$, and
the last term stands for an energy dependent incoherent pump with strength
$P_e$. The energy dependence is encoded in the generalized operators
$\tilde{p}_{j,\scriptscriptstyle I} = \sum_{n,\alpha,\alpha'}
S_{n-1\alpha',n\alpha} \braket{n-1,\alpha'| p_{j,\scriptscriptstyle I}
|n,\alpha} \ket{n-1,\alpha'} \bra{n,\alpha}$, which are expressed in the
eigenbasis $H \ket{n,\alpha} = \omega_{n,\alpha}\ket{n,\alpha}$ with
eigenenergies $\omega_{n,\alpha}$. Here, the index $\alpha$ labels the states
for a given photon number $n$.  The jump operators $\tilde{p}_{j,
\scriptscriptstyle I}$ induce transitions between different eigenstates of $H$
with a probability determined by the spectral weight of the reservoir
\begin{equation}
   S_{n-1\alpha',n\alpha} = \frac{\Gamma/2}{-i(\omega_{n,\alpha} -
   \omega_{n-1,\alpha'} - \omega_s) + \Gamma/2}
\label{spectral_weight} 
\end{equation}
The latter is Lorentzian-shaped with a width $\Gamma$ (see Fig.~\ref{fig2}(a)).
Note, that for a white noise bath ($\Gamma\rightarrow\infty$) one recovers a
standard Lindblad term with $\tilde{p}_j \rightarrow p_j$.

We now show that one can engineer a quasi-pure steady-state by carefully
designing the parameters of the pump.  In particular, it is the spectral
weight $S_{n-1\alpha',n\alpha}$ which determines the eigenstates that are
mostly populated by the drive. Centering the spectral profile
\eqref{spectral_weight} at zero energy, i.e., $\omega_s = 0$, implies that as
long as the width $\Gamma$ and the strength $P_e$ are sufficiently small
($\Gamma,P_e < \Delta$), only the (flat) zero-energy manifold is effectively
pumped by the reservoir. In this case, we project the density matrix on the
flat band manifold selected by the projector $\mathcal{P} =
\ket{\text{vac}}\bra{\text{vac}} + \dots + \ket{n_\text{dw}}\bra{n_\text{dw}}$
and $\rho \approx \mathcal{P}\rho \mathcal{P}$. In this flat band
eigenspace, we solve the master equation \eqref{me} analytically (see
SM). The density matrix then is a diagonal mixture of flat band
eigenstates with occupation probabilities given by
\begin{equation}
   \frac{\rho_n}{\rho_{n_\text{dw}}} 
   =  D_n \left(\frac{\kappa}{P_e}\right)^{n_{\rm dw} - n}
\label{me_analytic}
\end{equation}
where $\rho_n$ is the population of the manifold with $n$ photons in the flat
band, $\rho_n = \sum_{\alpha} \rho_{n,\alpha}$, $\rho_{n,\alpha} =
\braket{n,\alpha| \rho | n,\alpha}$ with $n=0,\dots,n_{\rm dw}$.
Consequently, when $P_e \gg \kappa$, we find $\rho \approx \ket{n_{\rm
dw}}\bra{n_{\rm dw}} + \mathcal{O}{(\kappa/P_e)}$, i.e., a quasi-pure
density-wave state.

In order to check the validity of this analytic result, we solve the
master equation numerically in the full lattice Hilbert space using a block
diagonalization algorithm (see SM).  We consider a finite-size system
with $N=5$ unit cells, i.e., $N_{\rm s} = 17$ sites, corresponding to a
density-wave with $n_{\rm dw} = 3$. Figs.~\ref{fig2}(c,d) show the
population of the density-wave state $\rho_{n_\text{dw}}$ in a colour-scale
map as a function of the Lorentzian width $\Gamma/\Delta$ and pump strength
$P_e/\kappa$. Indeed, we obtain an almost pure density wave state if two
conditions are fulfilled: (i) The spectral width and the strength of the
reservoir are smaller than the excitation gap ($\Gamma, P_e < \Delta$)
such that only flat band states are effectively excited; (ii) the pump
strength is larger than the dissipation rate ($P_e\gg \kappa$) such
that the density wave state has the largest population among all flat band
states.

\begin{figure}[t!]
\includegraphics[width=0.425\textwidth]{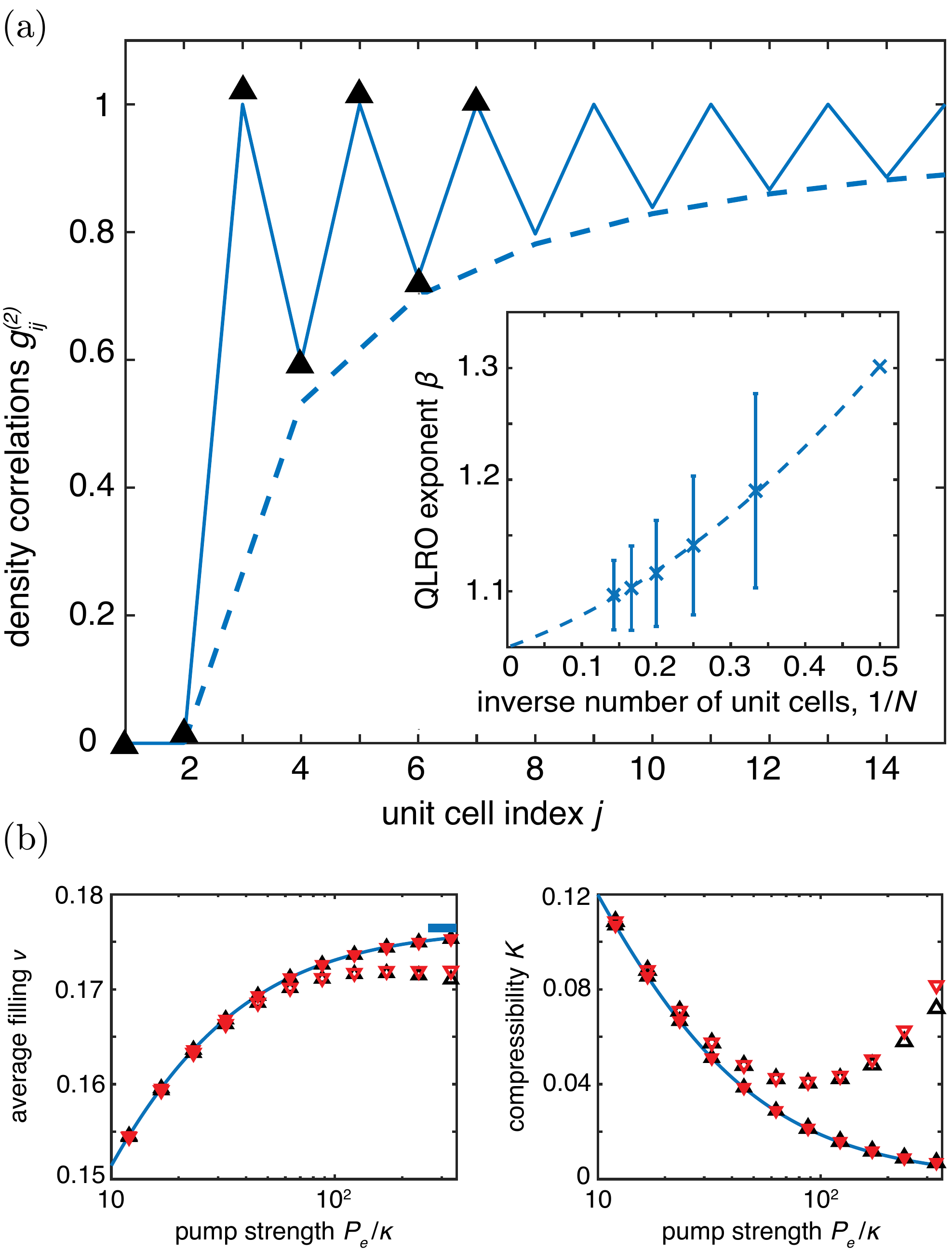}
\caption{(a) Density-density correlator of $C$ sites $g^{\scriptscriptstyle (2)}_{i,j}$
showing quasi-long-range order (QLRO).  The result obtained from the
projected Hilbert space for a lattice with $N=15$ unit cells (line) agrees
well with the exact numerics for a smaller system of $N=7$ unit cells
(symbols). The pump strength is $P_e/\kappa\!=\!75$ and we have chosen
$\Gamma/\Delta \!=\! 1.5$ in our numerical analysis, with
$\Delta/J\!  \approx\!  3\!\times\!10^{-4}$ the excitation gap for $N\!=\!7$.
The fitted envelope of the density-wave oscillations (dashed) decays
algebraically with an exponent $\beta$ (inset). (b) Average filling
$\nu=\bar{n}/N_{\rm s}$ (left) and compressibility $K=(\sum_n n^2\rho_n -
\bar{n}^2)/\bar{n}$ (right) as a function of pump strength $P_e/\kappa$ for
fixed values of the Lorentzian width $\Gamma/\Delta = 0.13, 1.3$.  Here,
$\bar{n} = \sum_n n \rho_n$ denotes the population of the manifold with photon
number $n$ and $\rho_n = \sum_{\alpha} \rho_{n,\alpha}$. The analytical
solution of the master equation \eqref{me} in the projected Hilbert space
\eqref{me_analytic} (lines) agrees well with the exact numerics for small
values of the Lorentzian width $\Gamma/\Delta = 0.13$ (filled symbols, on top of the line).
Deviations occur when the spectral width $\Gamma/\Delta=1.3$ (open symbols) 
is comparable with the excitation gap (see Fig.~\ref{fig2}) at large pump strengths. 
The black upward (red downward) triangles are obtained
including up to $n\!=\!n_{\rm dw}\!+\!1$ ($n\!=\!n_{\rm dw}\!+\!2$)
excitations in the full lattice Hilbert space, with good convergence obtained
in the data.
\hfill \label{fig3}}
\end{figure}

It is the analysis of the spatial correlations which brings forward the key
properties of this interesting nonequilibrium steady-state. We
determine the density-density correlator (second-order coherence of the
electromagnetic field) emitted by the $C$ sites in the lattice, i.e.,
$g^{\scriptscriptstyle (2)}_{i,j} = \langle p_{i,\scriptscriptstyle C}^\dagger
p_{j,\scriptscriptstyle C}^ \dagger p_{i,\scriptscriptstyle C}
p_{j,\scriptscriptstyle C}\rangle/(\langle p_{i,\scriptscriptstyle C}^\dagger
p_{i,\scriptscriptstyle C}\rangle\langle p_{j,\scriptscriptstyle C}^\dagger
p_{j,\scriptscriptstyle C}\rangle)$.  Fig.~\ref{fig3}(a) shows the spatial
correlations of the first $C$ site ($i=1$) with its neighbors as obtained from
the solution of the master equation in the projected Hilbert space (line) for
a lattice with $N=15$ unit cells. Remarkably, we find extended density-wave oscillations
whose envelope (dashed) decays algebraically,
\begin{equation}
   g^{\scriptscriptstyle (2)}_{1,2j} = 1 - 1/j^\beta 
   + \mathcal{O}[(\kappa/P_e)^2]
\label{eq:correlator}
\end{equation}
Equation \eqref{eq:correlator} marks the appearance of double-periodic
quasi-long-range order in this system and is the central result of our paper,
i.e., a truly crystalline state of light, induced by frustration and 
pump engineering. Making use of a log-log linear-regression fit and
extrapolating to $N\rightarrow\infty$ (inset in Fig.~\ref{fig3}(a)), we find
$\beta = 1.05\pm0.02$.  This estimate is consistent with the analytic solution
to first order in $\kappa/P_e$ (see SM), which yields $\beta=1$. 
This result is also compared with exact simulations for a smaller lattice with $N=7$ unit cells and excellent
agreement is obtained. The crystalline state of light is further characterized by a geometric filling
factor of $\nu=n_{\rm dw}/N_{\rm s}\approx 1/6$ and vanishing compressibility
as shown in Fig.~\ref{fig3}(b) (see Figure caption for a more detailed
discussion). We note that a different mechanism 
for achieving long-range crystalline order based on the combination of electromagnetically induced transparency 
and dipolar interactions between Rydberg polaritons was also proposed in \cite{chang2008,otterbach2013}.

We now provide an estimate of the parameters for a proof-of-principle
implementation using circuit QED, i.e., nonlinear microwave resonators, which
are coupled capacitively to each other \cite{schmidt2013*2}. In the dispersive
regime of circuit QED \cite{hoffman2011}, a Kerr nonlinearity $U\!\approx\!
2\,$MHz, inter-cavity coupling strength $J\!\approx\!200\,$MHz and a cavity
decay rate $\kappa\!\approx\!5\,$kHz are readily achievable with current
technology \cite{schmidt2013*2}. In order to engineer one period of a
density-wave, we need at least $N\!=\!4$ unit cells with $n_{\rm dw} = 2$. For
this system size and parameters, the minimal excitation gap would be
$\Delta\!\approx\!0.001J\!\approx\!2\,$MHz (calculated as in
Fig.~\ref{fig2}b).  In order to stay in the small bandwidth ($\Gamma, P_e <
\Delta$) and strong pump regime ($P_e \gg \kappa$), we can choose $\Gamma\sim
1\,$MHz and $P_e \sim 100\,$kHz.  Going to extended arrays of superconducting
qubits with stronger Kerr nonlinearities would allow for a large-scale
realization of our proposal \cite{neill2017}. An alternative platform for
the implementation of the photonic density-wave are exciton-polaritons in
semiconducting microcavities, where a Lieb chain has already been engineered
\cite{baboux2016}.

An interesting question for future work concerns the strong pump regime,
where dispersive states become significantly occupied and density-wave
oscillations are expected to vanish. In equilibrium, the density-wave state 
is destroyed by particle doping in the quasi-1D sawtooth chain but survives 
together with superfluidity in the 2D kagome lattice giving rise to a supersolid phase \cite{huber2010}. 
Whether such a picture remains valid out of equilibrium is an open question, 
with the potential for the discovery of novel photonic phases. 
An equally interesting direction concerns the nonequilibrium 
many-body system in the presence of disorder, given that 
the single-particle eigenstates become critical when 
the dispersive band and the flat band touch, e.g., 
in the 2D Lieb or kagome lattice \cite{chalker2010,leykam2013}.

\section*{Supplemental Material} 
\paragraph*{Flat band projection.} The flat band manifold is given by the set
of all noninteracting (non-overlapping) flat band states that can be written
starting from the single-particle plaquette states \eqref{lambda_state}, i.e.,
the single excitation (photon) states $\ket{1,\alpha} =
\ket{V_1},\ket{V_2}$ etc., the two-photon states $\ket{2,\alpha} =
\ket{V_1} \ket{V_3}, \ket{V_1} \ket{V_4}$ etc., the
three-photon states $\ket{3,\alpha} = \ket{V_1} \ket{V_3}
\ket{V_5}, \ket{V_1} \ket{V_3} \ket{V_6}$ etc., up to
the maximally excited density-wave state $\ket{n_{\rm dw}}$ in
\eqref{dw_state}, with the addition of the vacuum state $\ket{0} =
\ket{\text{vac}}$ as well. Here, $\alpha$ labels the flat band eigenstates
within each excitation number sector $n$ with degeneracy $D_n = \binom{2
n_{\rm dw} - n}{n}$. Note that the sectors with $n=0,~n_{\rm
dw}$ are special with $D_0 = D_{n_{\rm dw}} = 1$. The entire set forms the
macroscopically degenerate zero-energy manifold with degeneracy $F_{2n_{\rm
dw} + 1}$ discussed in the main text.  Note, that the single-particle
plaquette states \eqref{lambda_state} are linearly independent but not
orthogonal since neighboring plaquettes share one site.  We thus perform
(numerically) a Gram-Schmidt orthogonalization of the $\ket{n,\alpha}$ for
each $n$ to obtain the orthogonalized new basis $\ket{n,\alpha}_{o}$.  The
orthogonalization procedure allows to construct the projector 
\begin{equation}
   \mathcal{P} = \sum_{n=0}^{n_{\rm dw}}\sum_{\alpha=1}^{D_n}
   \ket{n,\alpha}_{o\,o}\!\bra{n,\alpha}
\label{methods_fbproj}
\end{equation}
which we use to project the density matrix $\mathcal{P}\rho\mathcal{P}$ and
the operators $\mathcal{P}p_j\mathcal{P}$.  The master equation then
reduces to a rate equation for the populations of the flat band states (the
diagonal entries of the density matrix), which can be solved analytically, see
equation \eqref{eq:sys_fb} below.  The solution for the populations in
\eqref{me_analytic} does not depend on the orthogonalization, while the
correlator $g^{2}_{1,j}$ in \eqref{eq:correlator} does, since it is
determined by the spatial structure of the eigenstates.

\paragraph*{Exact diagonalization.} 
We diagonalize the Hamiltonian \eqref{hammodel} in each excitation number
sector separately to obtain a complete basis $\ket{n,\alpha}$,
$H\ket{n,\alpha} = \omega_{n,\alpha}\ket{n, \alpha}$. Here, $\alpha$ labels
all states in the excitation number sector $n$, see e.g. Fig.~\ref{fig2}(a).  We
then write the master equation \eqref{me} in this eigenbasis and obtain a rate
equation for the populations $\rho_{n,\alpha} \!=\!
\braket{n,\alpha|\rho|n,\alpha}$, i.e.,
\begin{equation}
\begin{split}
   \dot{\rho}_{n,\alpha} = & \sum_{\alpha'}\Big[\kappa G_{n,\alpha,\alpha'}
   \rho_{n+1,\alpha'} + P_e \tilde{G}_{n,\alpha\alpha'}\rho_{n-1,\alpha'}\Big]\\
   & - \rho_{n,\alpha}(\kappa\,{W}_{n,\alpha} + P_e\,\tilde{W}_{n,\alpha})
\label{eq:rate}
\end{split}
\end{equation}
with $G_{n,\alpha,\alpha'}\!\!=\!\! \sum_{j,I}\!\braket{n,\alpha|
p_{j,\scriptscriptstyle I} | n\!+\!1,\alpha'}\!|^2$,
$\tilde{G}_{n,\alpha,\alpha'} \!=\!
\text{Re}[S_{n\alpha,n\!-\!1\alpha'}]\sum_{j,I}\!|\!\braket{n,\alpha|
p^\dagger_{j,\scriptscriptstyle I} | n-1,\alpha'}\!|^2$,
$W_{n,\alpha}=\sum_{j,I}\braket{n,\alpha | p^\dagger_{j,\scriptscriptstyle
I}p_{j,\scriptscriptstyle I} | n,\alpha}$ and $\tilde{W}_{n,\alpha} \!=\!
\sum_{\alpha'} \tilde{G}_{n+1,\alpha',\alpha}$. 
The first two terms in equation \eqref{eq:rate} describe the incoherent
excitation of level $n$ due to losses from level $n+1$ and pump from level
$n-1$, while the last two terms describe the corresponding losses out of the
sector $n$.  Solving the rate equations allows to circumvent the computational
costs of the master equation, since we only need to obtain the diagonal
entries of $\rho$.  This approach allows to analyze lattice sizes of the order
of 20 sites, i.e., Hilbert spaces with $>10^6$ states. Such system sizes are
typically far beyond exact diagonalization methods (note, that tensor network
approaches are not directly applicable here since a matrix representation of
the generalized operators $\tilde{p}_j$ requires to first solve for the
complete eigenbasis of $H$, i.e., including the excited states). The rate
equation \eqref{eq:rate} is solved numerically imposing a
cutoff in the number of photons per site $n_{\rm loc}$ and a global cutoff in
the number of excitations in the lattice $n_{\rm glob}$. We then verify
convergence in these truncation parameters. All converged 
results shown are obtained using up to $n_{\rm loc}=2$ and $n_{\rm glob} = 5$.

\paragraph*{Rate equation in the projected Hilbert space.} 
For flat band states, the rate equation \eqref{eq:rate} simplifies greatly,
since $S_{n-1\alpha',n\alpha} = 1$, $\sum_\alpha G_{n,\alpha,\alpha'} = n+1$,
$\sum_{\alpha'}\tilde{G}_{n,\alpha,\alpha'} = n$, $W_{n,\alpha}=n$ and $\sum_\alpha \tilde{W}_{n,\alpha} = (n+1)D_{n+1}$. Summing
over $\alpha$ in equation \eqref{eq:rate}, one finds
\begin{equation}
\begin{split}
   \dot{\rho}_{n} & = \kappa (n+1)\rho_{n+1} 
   + P_e n\frac{D_n}{D_{n-1}}\rho_{n-1} \\
   & -  \kappa n \rho_{n}  - P_e (n+1)\frac{D_{n+1}}{D_n}\rho_{n}
\label{eq:sys_fb}
\end{split}
\end{equation}
with $\rho_n = \sum_{\alpha}\rho_{n,\alpha}$ and $n=0,\dots,n_{\rm
dw}$. The solution is easily determined analytically,
\begin{equation}
\label{eq:sol_methods}
\frac{\rho_{n+1}}{\rho_{n}} = \frac{P_e}{\kappa}\frac{D_{n+1}}{D_n}
\end{equation} 
from which we obtain the expression \eqref{me_analytic}.
\vspace{5pt}
\paragraph*{Density-density correlator in the projected Hilbert space.} 
In the strong pumping regime $P_e\gg \kappa$, the density matrix takes
the form $\rho \approx \ket{n_{\rm dw}}\bra{n_{\rm dw}}  +
(\kappa/P_e)\sum_{\alpha \in D_{n_{\rm dw} - 1}} \ket{n,\alpha}\bra{n,\alpha}
+ \mathcal{O}[((\kappa/P_e)^2]$. The correlator $g^{\scriptscriptstyle
(2)}_{i,j} = \langle p_{i,\scriptscriptstyle C}^\dagger
p_{j,\scriptscriptstyle C}^ \dagger p_{i,\scriptscriptstyle C}
p_{j,\scriptscriptstyle C}\rangle/(\langle p_{i,\scriptscriptstyle C}^\dagger
p_{i,\scriptscriptstyle C}\rangle\langle p_{j,\scriptscriptstyle C}^\dagger
p_{j,\scriptscriptstyle C}\rangle)$ is calculated using the formula $\langle O
\rangle  = \text{Tr}[\rho O]/\text{Tr}[\rho]$. It is straightforward to show
that $g^{\scriptscriptstyle (2)}_{1,2j+1} = 1 + \mathcal{O}[(\kappa/P_e)^2]$
since the expectation values are dominated by the density-wave contribution to
the density matrix. Here, $j=1,\dots,(N-1)/2$, while $g^{\scriptscriptstyle
(2)}_{1,1}=0$ trivially. Calculation of the correlator $g^{\scriptscriptstyle
(2)}_{1,2j}$, $j=1,\dots,(N-1)/2$ is more involved and the contribution of the
manifold below the density-wave, i.e., with $n=n_{\rm dw} - 1$, must be taken
into account. Neglecting the orthogonalization of the states in this
mani\-fold, we find $g^{\scriptscriptstyle (2)}_{1,2j} = 1 - 1/j +
\mathcal{O}[(\kappa/P_e)^2]$.


\vspace{10pt}
\section*{Acknowledgements}
 
We thank I.\ Carusotto, E.P.L.\ v. Nieuwenburg and H.E.\ T\"ureci for
discussions and acknowledge support from the Swiss National Science Foundation
through the National Centre of Competence in Research `QSIT--Quantum Science
and Technology' (MB).

\end{document}